\documentclass[nohyperref]{article}

\usepackage{microtype}
\usepackage{graphicx}
\usepackage{subfigure}
\usepackage[titletoc,title]{appendix}
\usepackage{booktabs} 

\usepackage{hyperref}

\usepackage{PRIMEarxiv}

\usepackage{amsmath}
\usepackage{amssymb}
\usepackage{mathtools}
\usepackage{amsthm}

\usepackage[capitalize,noabbrev]{cleveref}

\theoremstyle{plain}

\theoremstyle{definition}

\theoremstyle{remark}

\usepackage[textsize=tiny]{todonotes}

\usepackage{amsmath}
\usepackage{xcolor}
\usepackage{natbib}

\usepackage[utf8]{inputenc}

\title{\textsc{Validating Causal Inference Methods}}
\author{
  Harsh Parikh\\
  Duke University \\
  \texttt{harsh.parikh@duke.edu}
  \And 
  Carlos Varjao\\
  Amazon.com \\
  \texttt{varjaoc@amazon.com}
  \And
  Louise Xu\\
  Amazon.com \\
  \texttt{louisexu@amazon.com}
  \And 
  Eric Tchetgen Tchetgen \\
  University of Pennsylvania \\
  \texttt{ett@wharton.upenn.edu}
  }

\begin{document}

\maketitle
\begin{abstract}
The fundamental challenge of drawing causal inference is that counterfactual outcomes are not fully observed for any unit. Furthermore, in observational studies, treatment assignment is likely to be confounded. Many statistical methods have emerged for causal inference under unconfoundedness conditions given pre-treatment covariates, including: propensity score-based methods,  prognostic score-based methods, and doubly robust methods. Unfortunately for applied researchers, there is no `one-size-fits-all' causal method that can perform optimally universally.In practice, causal methods are primarily evaluated quantitatively on handcrafted simulated data. Such data-generative procedures can be of limited value because they are typically stylized models of reality. They are simplified for tractability and lack the complexities of real-world data. For applied researchers, it is critical to understand how well a method performs for data at hand. Our work introduces a deep generative model-based framework, \textit{Credence}, to validate causal inference methods. The framework's novelty stems from its ability to generate synthetic data anchored at the empirical distribution for the observed sample, and therefore virtually indistinguishable from the latter. The approach allows the user to specify ground truth for the form and magnitude of causal effects and confounding bias as functions of covariates. Thus simulated data sets are used to evaluate the potential performance of various causal estimation methods when applied to data similar to the observed sample. We demonstrate Credence's ability to accurately assess the relative performance of causal estimation techniques in an extensive simulation study and two real-world data applications from Lalonde and Project STAR studies.
\end{abstract}

\section{Introduction}
Causal inference problems typically focus on estimating the effect of a point treatment ($Z$) on an outcome of interest ($Y$) \cite{angrist2008mostly} - one example would be determining if a drug will aid in a patient's recovery. In a binary treatment setting, the average causal effect ($\tau$) of an intervention is defined as the contrast between the population average potential outcome if all units were given the active treatment ($\mathbf{E}[Y(1)]$) versus the average potential outcome if  all units were given the control treatment ($\mathbf{E}[Y(0)]$). In this paper, the contrast of interest is defined on the additive scale, that is $\tau = \mathbf{E}\left[ Y(1) - Y(0) \right]$, which hereafter is referred to as the average treatment effect or ATE.

We will proceed under the standard consistency assumption typically made in the causal inference literature, that for each unit with observed treatment value, say  $Z=z$, its observed outcome matches the potential outcome $Y(z)$ while the counterfactual outcome $Y(1-z)$ remains unobserved or \textit{missing}. 
Hence, knowing the true causal effect of intervention for each individual is impossible, however, it is possible under certain conditions to learn the population average causal effect  $\tau$. A standard approach to identify the ATE relies on an assumption that one can account for the possibility of confounding, on the basis of measure covariates, therefore ruling out the presence of unmeasured confounding. Confounding factors are typically common causes of the treatment and outcome variables, for instance, patients that were able to avail themselves of a treatment drug might be affluent and have better access to healthcare facilities compared to patients in the control group. Thus, the correlation between the treatment and the outcome might be primarily because of better access to healthcare. In order to evaluate the ATE,  variables such as ``access to healthcare'' in this case, that simultaneously affect the treatment ($Z$) and the outcome ($Y$) are confounders ($W$) that must be accounted for. In addition  to consistency and unconfoundedness, nonparametric identification of the ATE typically also requires a positivity assumption that for any covariate value observed in the population, there is at least one unit who would receive the active treatment and at least one unit that would receive the control treatment.

\paragraph{Causal effect estimation. } 
Under these identifying conditions, the causal inference literature has over the years produced several alternative approaches to  explicitly account for observed covariates with the goal to adjust for confounding. This rich analytical arsenal includes three types of methods distinct in the manner in which covariate adjustment is achieved: (i) methods that directly model the dependence of the outcome variable on covariates, such as nearest neighbor matching or outcome regression adjustment; (ii) methods that model dependence of the treatment variable on covariates (also known as treatment propensity score), such as propensity score matching or inverse-probability-of-treatment-weighting; and (iii) so-called doubly robust methods that combine both approaches for improved robustness  \cite{rosenbaum1984reducing,imbens2015causal, hernan2010causal, van2003unified, van2018targeted}.
Conditions under which methods (i)-(iii) 
 can provide valid inferences about ATE are well understood, and typically require that bias incurred by the approach in estimating nuisance parameters is sufficiently small. Such bias will be small if nuisance functions of covariates, e.g the conditional mean for the outcome in the control group as a function of covariates or the propensity score, are smooth enough to be estimated at sufficiently fast rates.
 However, performance of these methods might vary significantly for any given finite size data-sample.  For instance, for the well-known Lalonde's National Support Work Demonstration (NSW) temporary employment program and income dataset \cite{Lalonde} where the treatment is randomly assigned, the average treatment effect (ATE) estimates using different methods are vastly different (see Table~\ref{tab: nsw}). Each causal inference estimation method has both strengths and limitations, and the optimal choice of estimation can depend on the problem at hand, in the sense that the best method in one application may be sub-optimal for the next application depending on the underlying data generation mechanism. For instance, parametric methods such as those based on linear regression work well even with limited data if the assumed model(s) are in congruence with the underlying true mechanism of dependency \cite{angrist2008mostly}. However, under model misspecification, the performance of these methods significantly deteriorates. Matching-based approaches do not have to explicitly assume an outcome model, however, it has been shown that the result can be highly sensitive to the choice of distance metrics, whether the features being matched are scaled appropriately or other hyper-parameters \cite{Stuart10,diamond2013genetic,malts,flame}. Non-parametric methods such as double machine learning with gradient boosting trees and targeted maximum likelihood estimation (TMLE) with superlearner can require relatively large datasets and fine-tuning of hyper parameters to ensure optimal performance \cite{athey2019machine, yifan2019selective, van2007super, cvtmle}. Thus, there is clearly no `\textit{one-size-fits-all}' estimation method for all scenarios. Validation of causal estimation methods for a given application is an important and challenging area of great consequence for causal inference practitioners, which to-date remains under-developed despite having received some attention.  

\begin{table}
    \centering
        \caption{Lalonde's NSW Sample ATE Estimated using few commonly used causal effect estimation methods.}
    \label{tab: nsw}
    \footnotesize
\begin{tabular}{lrr}
\\\toprule
{\textbf{Estimators}} &  \textbf{ATE Estimate} &  \textbf{Std. Dev.} \\
\midrule
Difference of Means       &   886.30 &  277.37 \\
Double Machine Learning   &   370.94 &  394.68 \\
Causal BART               &   818.79 &  184.46 \\
Propensity Score Matching &  1079.13 &  158.59 \\
\bottomrule
\end{tabular}
\vspace{-5mm}
\end{table}

\paragraph{Existing approaches to validating causal methods.}
 Existing approaches in the literature to validate or evaluate the performance of causal methods on a specific dataset of interest can be divided into three main categories: (1) face-validity test, (2) placebo or negative control tests, and (3) synthetic data tests. Face-validity tests assess the estimated treatment effect against an expert’s intuition. For instance, if the expected effect of a malaria medication according to a pharmacologist was a reduction in risk of death, but a causal method’s estimate was contrary to the belief, then this fails the face-validity test. This test clearly is neither sufficient nor necessary - violation does not necessarily indicate that the causal estimation method is biased while agreement with expert’s intuition does not imply that the causal estimation method is accurate. Placebo tests are the most widely used validation approach for causal models and they come in two varieties: “in-time” and “in-sample” placebo tests \cite{abadie2010synthetic,ferman2017placebo,athey2017state}. An in-time-placebo test either restricts the analysis to a time-period whereby the causal effect is known to be null, or reassigns the treatment to a time period before the actual observed treatment and estimates the treatment effect to generate the distribution of treatment effects under null. In-sample-placebo tests, on the contrary, reassign the treatment to a control unit unaffected by the treatment, and use the causal inference method to estimate the effect.  These tests are valid only under strong assumptions such as: 
 (a) for in-time placebo tests, selection into treatment is assumed to be independent of pre-treatment outcomes which can lead to reverse causal-dependency bias,
 (b) in-sample placebo tests assumes that the placebo treatment assignment preserves key features of the true treatment assignment mechanism.
 Closely related to (a) and (b), negative control exposure or outcome variables are also routinely used for causal validation \cite{lipsitch2010negative, shi2020selective}
 Such methods require  that negative control exposure and negative control outcome variables are available such that they satisfy key exclusion restrictions and U-comparability conditions \cite{lipsitch2010negative}. 
 
Lastly, for synthetic data tests, the design typically includes a known data-generating mechanism. This allows access to treatment effects ground truth used to evaluate the accuracy of causal methods on simulated data. However, synthetic data generating processes are commonly handcrafted, oversimplified, and thus unlikely to reflect the same level of complexity as real-world data. In addition, these overly-stylized data generating processes tend to a priori favor certain methods over others \cite{athey2021using,advani2019mostly,knaus2021machine,gentzel2019case,gentzel2021and}. Hence, good performance of a given method on handcrafted synthetic data does not necessarily translate to equally good performance on real-world data. In this paper,  
we present an approach to generate synthetic data sets that satisfy two salient properties sought out in simulation studies: (i) user-specified causal treatment effects, heterogeneity, and endogeneity; (ii) simulated samples that are stochastically indistinguishable from the observed data sample of interest.

\paragraph{Contribution. } 
Specifically, we introduce a general framework (\textit{Credence}) to validate and evaluate the performance of various existing causal inference methods using synthetic data anchored at the empirical distribution of a given data set of interest (see Section~\ref{sec:framework}). The approach relies on a deep generative model trained on a rich universe of data sets that share certain key features with the data set of primary interest, and is the basis to operationalize the proposed framework (see Section~\ref{sec:Credence}). \textit{Credence} successfully fulfils requirements (i) and (ii), central to obtaining an objective evaluation of causal methods to a data set in view. 

A recent works by \citet{athey2021using} and \citet{brady2020} similarly proposed to use of deep generative models to simulate synthetic data to validate causal methods. However, in contrast to our approach, \citet{athey2021using} assumed that the observed data satisfied conditional ignorability, a restriction which they also imposed in generated samples. 
Furthermore, both their proposed framework does not appear to accommodate user-specified treatment effects and confounding bias. As result, their approach only fulfils requirement (ii) but not (i), a important gap we address in this work.

We demonstrate \textit{Credence}'s utility  with two synthetic data scenarios (read Section~\ref{sec:synthdata}), and on two real data examples -- the Lalonde and Project STAR studies (read Section~\ref{sec:realdata}). For synthetic data experiments, we compare the oracle performance ranking of various causal inference estimation methods in a Monte Carlo setting where the true data-generating process (DGP) is completely known and compare it with the performance evaluation produced by  \textit{Credence} without a priori knowledge of the DGP. We find that \textit{Credence} can successfully re-produce the oracle performance ranking of the various estimation methods based on the single observed sample. Furthermore, in each real-world data set for which both experimental and observational data are available, we compare the performance ranking for the same set of causal inference methods in the  observational sample with that in the experimental sample. Finally, we discuss important implications of \textit{Credence} and highlight both strengths and limitations of the proposed framework in  Section~\ref{sec:discuss}.
\section{Methodology}\label{sec:method}
In this section, first we introduce the general framework (Section~\ref{sec:framework}) to learn a data generating process from the observed data set of interest, with user-defined levels of treatment effects and confounding. Next, we introduce \textit{Credence}, which operationalizes the proposed framework using an ensemble of variational autoencoders (Section~\ref{sec:Credence}).
\subsection{Framework}\label{sec:framework}
Intuitively, our first objective is to learn a data generating mechanism, $G_{\theta}$, such that the joint distribution of the variables in a generated sample $D' = (X',Y',Z') \sim G_{\theta}$ is nearly indistinguishable from the empirical distribution of the observed data $D = (X,Y,Z)$. Crucially, in addition to the above feature, we incorporate user-specified causal treatment effects, heterogeneity and both observed and hidden confounding bias as constraints that must be satisfied to the extent possible by the learned law $G_{\theta}$ and therefore can stand as a causal ground truth anchor for generated samples.

\begin{figure}
    \centering
    \includegraphics[width=0.8\linewidth]{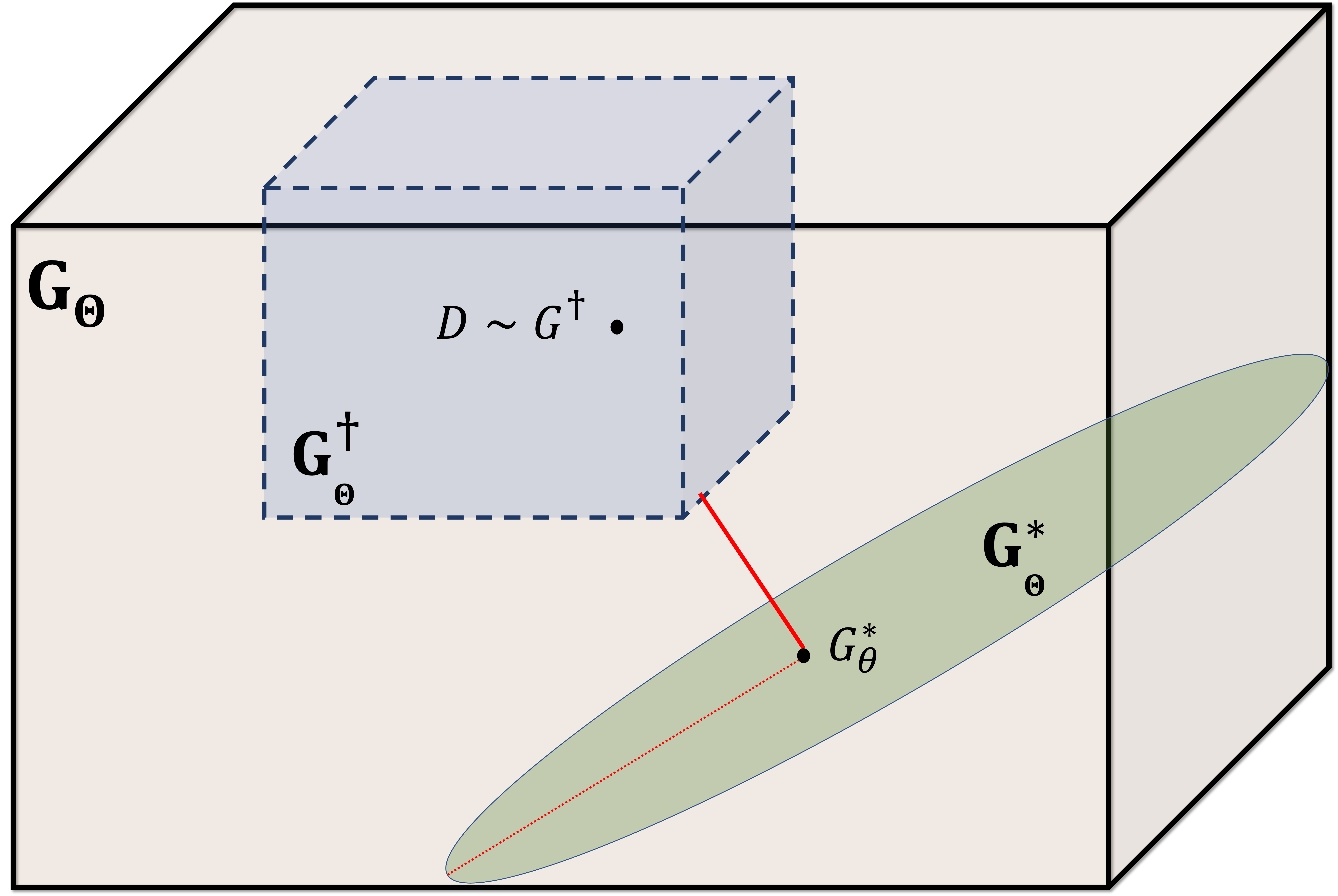}
    \caption{Schematic drawing of the optimization setup. Here $\mathbf{G}_\Theta$ is the space of all possible data generators, $G^\dagger$ is the true data generative process for the observed data $D$, $\mathbf{G}^\dagger_\Theta$ is the subspace of all equally likely data generators for observed data $D$, and $\mathbf{G}^*_\Theta$ is the subspace of potential data generators that can be learned using the assumed configuration. We want to find a $G^*_\theta \in \mathbf{G}^*_\Theta$ that is closest to a $G_\theta \in \mathbf{G}^\dagger_\Theta$.}
    \label{fig:schema_opt}
\end{figure}

In this vein, let $\mathbf{G}_\Theta$ denote a model consisting of possible data generating laws for $D$. Let $G^\dagger \in \mathbf{G}_\Theta$ denote the true generating model from which the observed data $D=(X,Y,Z)$ was sampled. For $\{\epsilon_{X,i},\epsilon_{Z,i},\epsilon_{Y,i}\}_{i=1}^N \overset{iid}{\sim} \mathcal{N}(\mathbf{0},\mathbf{1})$ and $\{U_i\} \overset{iid}{\sim} \mathcal{N}(0,1)$, throughout, we assume $D$ follows the nonparametric structural equations model: (1) $X_i \sim \phi_X(\epsilon_{X,i})$, (2) $Z_i \sim \phi_Z(X_i,U_i,\epsilon_{Z,i})$, (3) $Y_i(1) \sim \phi_{Y(1)}(X_i,U_i,\epsilon_{Y,i})$, (4) $Y_i(0) \sim \phi_{Y(0)}(X_i,U_i,\epsilon_{Y,i})$, and (5) $Y_i = Z_i Y_i(1) + (1-Z_i) Y_i(0)$ where, $G^\dagger = [\phi_X,\phi_Z,\phi_{Y(1)},\phi_{Y(0)}]$.
Further, let $\mathbf{G}^\dagger_\Theta \subseteq \mathbf{G}_\Theta$ be the subset of equally likely data generators for observed data $D$, and $\mathbf{G}^*_\Theta \subseteq \mathbf{G}_\Theta$ be the subset of feasible data generators that satisfy user-defined treatment effects and confounding bias  functions. Informally, if $\mathbf{G}^\dagger_\Theta \cap \mathbf{G}^*_\Theta \neq \{\emptyset\}$ then we wish to find an  arbitrary law $G^*_\theta \in \mathbf{G}^\dagger_\Theta \cap \mathbf{G}^*_\Theta$, otherwise, we wish to find a $G^*_\theta \in \mathbf{G}^*_\Theta$ which is ``closest'' to a $G_\theta \in \mathbf{G}^\dagger_\Theta$ (see Figure~\ref{fig:schema_opt} for a graphical illustration).
To formalize this idea, consider an optimization setup that takes in as input the observed data $D$; together with user-defined treatment effect function ($f$) and unmeasured confounding-bias function ($g$), both of which are defined as followed on the mean additive scale: 
\begin{eqnarray*}
f(x) &=& \mathbf{E}[Y(1) - Y(0)| X=x]\\
g(x,z) &=& \mathbf{E}[Y(z)|X=x,Z=z]\\ &&\;\; - \mathbf{E}[Y(z)|X=x,Z=1-z]
\end{eqnarray*}
The setup also accommodates specification of the degree to which the user wishes the data generating mechanism be faithful to specified functions $f$ and $g$, regulated by the pair of penalties, referred throughout as \textit{rigidness parameters} $\alpha$ for $f$ and $\beta$ for $g$, respectively. We design the optimization objective to finding the parameter $\theta$ of the generative model $G_\theta$ that minimizes a certain distributional distance metric between the empirical distribution of $D$ and the simulated data under specified treatment effect and confounding bias constraints. 

Specifically, one may consider the Wasserstein distance as choice of distributional distance metric; in which case, the treatment effect constraint shrinks the conditional average treatment effect towards $f(X)$ and the corresponding confounding bias constraint  shrinks the amount and nature of confouding bias towards $g(X,Z)$ in the fitted DGP. In case there is a trade-off between minimizing each of the three objectives, the rigidness coefficient is used to encode specific user's preferences for one over the other. More broadly, let $d$ denote the distance metric of choice between empirical and estimated laws; then we wish to learn a data generating mechanism $G_\theta$ that satisfies:
\begin{flalign}\label{eq:objective}
   &\textbf{min}_{\theta} \;\; \mathbf{E}\left[d( (X,Y,Z), (X',Y',Z') ) \right] \notag\\
& \qquad  + \alpha \left\| \mathbf{E}[Y'(1)-Y'(0)|X'=x'] - f(x') \right\|  \\
&  \qquad + \beta \left\|\begin{tabular}{l}
   $\mathbf{E}[Y'(z')|X'=x',Z'=z']$ \\
    $\;-\mathbf{E}[Y'(z')|X'=x',Z'=1-z']$ \\
    $ \;\;\;\;\;- g(x',z')$
\end{tabular}\right\| \notag
\end{flalign}
where $D'=(X',Y',Z') \sim G_\theta$.

Next, we discuss our procedure to operationalize this optimization using deep generative models.

\subsection{Credence}\label{sec:Credence}
\begin{figure}
    \centering
    \includegraphics[width=0.49\textwidth]{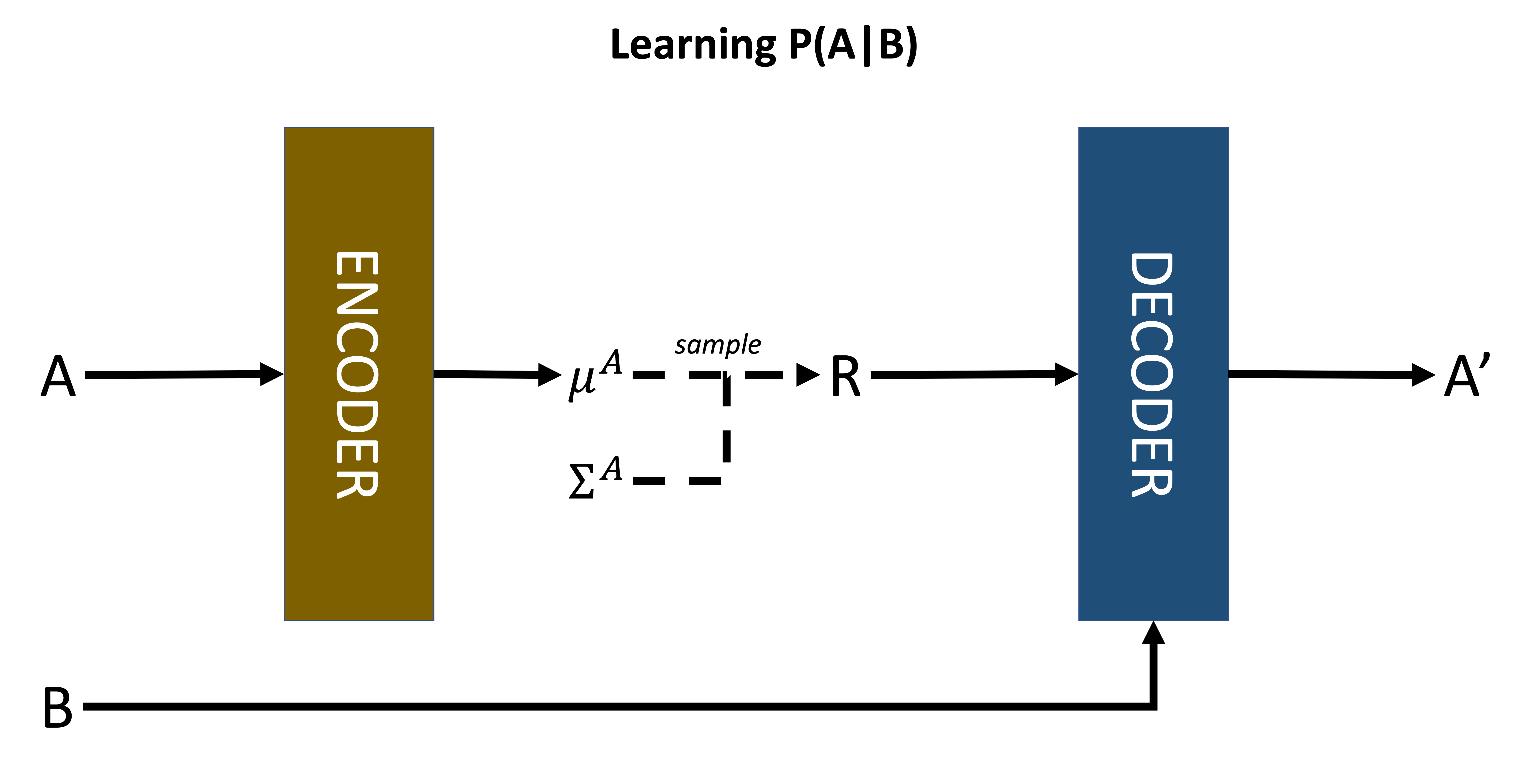}
    \caption{Schematic diagram showing conditional VAE (conVAE) used to learn a conditional distribution $P(A|B)$ for any two random variables $A$ and $B$.}
    \label{fig:conVAE}
\end{figure}

Credence uses this framework to generate data for comparing and validating causal inference methods. Credence learns the joint distribution $P(X,Z,Y)$
as an ensemble of three conditional distributions: $P(Y| X, Z)$, $P(X | Z)$ and $P(Z)$ using a variational autoencoder neural networks. A brief background on variational autoencoders (VAE) is discussed in Appendix~\ref{sec:vae}. Our work uses the VAE framework to learn the above mentioned conditional distributions by allowing the decoder to not only be a function of sampled latent vectors but also of variables we are conditioning on (see Figure~\ref{fig:conVAE}). In machine learning literature, VAEs have been shown to work extremely well for modeling numerical as well as vision and text data \cite{kingma2019introduction,kusner2017grammar}. However, to the best of our knowledge, deep generative models such as VAEs haven’t been used in the specific way described for validating and evaluating causal inference methods.

In this work, we have assumed $Z$ to be a binary indicator. Thus, learning a generative model for $P(Z)$ boils down to estimating $p_z=P(Z=1)$, the parameter of a Bernoulli distribution. Thus, the sample proportion $\widehat{p}_z = \frac{\sum_i Z_i }{N}$ is an unbiased and consistent estimate of $p_z$. We use the conditional VAEs to then learn probability distributions $P(X|Z)$ and $P(Y| X,Z)$ separately. 

Consider the VAE setup $H^{X|Z}_{(\phi_{X|Z},\theta_{X|Z})} = (F^{X|Z}_{\phi_{X|Z}},G^{X|Z}_{\theta_{X|Z}})$ to learn $P(X|Z)$. Here, $F^{X|Z}_{\phi_{X|Z}}$ is the encoder that maps the observed data to a lower dimensional latent space, and $G^{X|Z}_{\theta_{X|Z}}$ is the decoder that maps back points to the original dimensions. 
Thus, the VAE encoder maps $X_i$ to latent space as $[\mu^X_i,\Sigma^X_i] = F^{X|Z}_{\phi_{X|Z}}(X_i)$, then sample a point $R^X_i$ in latent space in the vicinity of $\mu^X_i$ by drawing it from the distribution $R^X_i \sim \mathcal{N}(\mu^X_i,\Sigma^X_i)$ and $R^X_i$ is mapped back to the space of $X$ using the decoder, $X'_i = G^{X|Z}_{\theta_{X|Z}}(R^X_i,Z_i)$.
We use the ELBO loss function for the setup which is extremely common in the VAE literature. It is given as follows,
\begin{eqnarray*}
    &&\mathcal{L}_{X|Z}(\phi_{X|Z},\theta_{X|Z}; \{(X_i,Z_i)\}_{i=1}^N) \\ &&=\sum_{i=1}^{N} 
        \| X_i - X_i' \|_2 + D_\textrm{KL}(\mathcal{N}(\mu^X_i,\Sigma^X_i) ||
        \mathcal{N}(0,1) ).
\end{eqnarray*}
Similarly, consider the VAE setup $H^{Y|X,Z}_{(\psi_{Y|X,Z},\theta_{Y|X,Z})} = (F^{Y|X,Z}_{\psi_{Y|X,Z}},G^{Y|X,Z}_{\theta_{Y|X,Z}})$ for learning $P((Y(1),Y(0))|X,Z)$. As noted before, for each unit in the original dataset we only observe one of $Y(0)$ or $Y(1)$. Thus, we update the loss function to inform the learning of $P((Y(1),Y(0))|X,Z)$ using the user-provided treatment effect function $f$ and confounding bias function $g$. That is, the constraints imposed by choices of $f$ and $g$ together with their corresponding penalties implicitly induce correlation between $Y(1)$ and $Y(0)$.  Let $[\mu^Y_i,\Sigma^Y_i] = F^{Y|X,Z}_{\psi_{Y|X,Z}}(Y_i)$, $R^Y_i \sim \mathcal{N}(\mu^Y_i,\Sigma^Y_i)$, $(Y_i'(1),Y_i'(0)) = G^{Y|X,Z}_{\theta_{Y|X,Z}}(R^Y_i,X_i,T_i)$ and $(Y_i''(1),Y_i''(0)) = G^{Y|X,Z}_{\theta_{Y|X,Z}}(R^Y_i,X_i,1-T_i)$. Then, the loss functions training $H^{Y|X,Z}_{(\psi_{Y|X,Z},\theta_{Y|X,Z})}$ is given by
\begin{eqnarray*}
    &&\mathcal{L}_{Y|X,Z}(\psi_{Y|X,Z},\theta_{Y|X,Z}; \{(Y_i,X_i,Z_i)\}_{i=1}^N) \\ &&=\frac{1}{N}\sum_{i=1}^{N} 
        \| Y_i - (Z_i Y_i'(1) + (1-Z_i) Y_i'(0)) \|_2 
    \\ && \qquad + D_\textrm{KL}(\mathcal{N}(\mu^Y_i,\Sigma^Y_i) ||
        \mathcal{N}(0,1) )
    \\ && \qquad + \alpha \| Y_i'(1) -Y_i'(0) - f(X_i) \|
    \\ && \qquad + \beta \| Y_i'(1-Z_i) - Y_i''(1-Z_i) - g(X_i,Z_i) \|
\end{eqnarray*}
where $\alpha$ and $\beta$ are the user-defined hyper-parameters. The expression $\mathcal{L}_{Y|X,Z}(\phi_{Y|X,Z},\theta_{Y|X,Z}; \{(Y_i,X_i,Z_i)\}_{i=1}^N)$ can be intuitively decoded -- the first and second term ensure distributional similarity of observed and generated data, while the second term also forces the distribution in the latent space to be standard normal. The third term shrinks the individualized treatment effects to the one defined by $f$, and the fourth term shrinks the magnitude and nature of confounding bias towards $g$. Note that the conditions enforced by the third and fourth terms of the loss function are tighter than the one described in equation~\ref{eq:objective}, thus narrowing the space of feasible generative models. This choice is to primarily ensures efficient computation and training. Also, note that if $\alpha=0$ then the treatment effects are free to assume any value, while if $\beta=0$ then the  unmeasured confounding can be arbitrary. 

This approach approximates the population-level optimization setup described in Section~\ref{sec:framework} and Equation~\ref{eq:objective}. Further, this setup allows each of the two VAEs to be trained independently. Hence, the procedure can be performed in parallel which reduces the overall time of analysis. 

\paragraph{Credence-based Validation.} Let $G^*_{\theta}$ be an optimal data generator that minimizes the objective described in Equation~\ref{eq:objective}. 
Then, we use the ensemble of decoders from the trained VAEs $\widehat{G}_{\theta} = \{ \widehat{G}_{\theta_{Y|X,Z}}, \widehat{G}_{\theta_{X|Z}}\}$ as an estimate of $G^*_{\theta}$. Further, $\widehat{G}_{\theta}$ is used to generate data $(X',Z',Y'(1),Y'(0))$ with known treatment effects such that $\forall j$, $R^{X'}_j\sim \mathcal{N}(0,1)$, $R^{Y'}_j\sim \mathcal{N}(0,1)$, $Z'_j \sim Bernoulli\left( \frac{\sum_i Z_i }{|D|} \right)$, $X'_j \sim \widehat{G}_{\theta_{X|Z}}(R^{X'}_j,Z'_j)$ and $(Y_j'(1),Y_j'(0)) \sim \widehat{G}_{\theta_{Y|X,Z}}(R^{Y'}_j,X'_j,Z'_j)$. 
The generated data is then used to assess the performance of various causal methods by comparing the estimated ATE to the true ATE, $E\{Y_j'(1)-Y_j'(0)\}$. 
In the remainder of the paper, we demonstrate Credence in two settings: (i) using synthetic data where ground truth is known; (ii) two real data applications each of which having available both observational and experimental samples. In settings (i) and (ii), Credence is trained on a universe of data sets generated via the nonparametric bootstrap from a single observed data sample. Credence is then used to conduct a Monte Carlo study anchored at a single realization constituting the observed sample, in order to compare the performance of the following set of canonical causal inference methods taken from published literature as representative of the landscape of available methods: (1) meta-learners \cite{metalearners}, (2) non-parametric double machine learning (DML) \cite{chernozhukov2018double} (3) propensity score matching estimator \cite{rosenbaum2002observational}, (4) causal forest \cite{wager2018estimation}, (5) targeted maximum-likelihood estimator (TMLE) \cite{van2006targeted}, (6) causal BART \cite{Chipman10bart:bayesian}, and (7) doubly robust estimator \cite{bang2005doubly}.
\section{Experiments}\label{sec:experiment}
In this section, we study \textit{Credence}-generated recommendations and rankings of causal inference methods for various datasets under different conditions. We divide this section into two parts. The first part focuses on synthetically generated data with known true data generating process (DGP)
. Here, we compare the performance of various causal inference methods on data sampled from the known true DGP. We compare these performances with the methods' performance on data sampled from three trained Credence models under different constraints. Ideally, we aim to evaluate the extent to which Credence can replicate the oracle's ranking under various sets of assumptions about treatment effects and confounding. The second part focuses on two real data case studies--(1) Lalonde data sets \citep{dehejia99}, and (2) Project STAR dataset \citep{project_star_data}. Both data sets have an experimental component (in which treatment is randomized) and an observational component (in which units self-select into treatment). We rank the performance of causal methods on the observational part of the data, using the difference of means estimate of ATE on experimental part as an unbiased estimator of ground truth. Further, we compare this ranking with the ranking produced using \textit{Credence} trained on the observational component of each study. 

\subsection{Synthetic Data Experiment}\label{sec:synthdata}
We use two DGPs  for our analysis. The first DGP generates normally distributed covariates with a quadratic treatment effect and selection into treatment  as a function of pre-treatment covariates. The second one is the so-called Friedman's DGP that which admits highly non-linear outcome and treatment selection functions of uniformly distributed covariates. Both the DGP's are described in details in Appendix~\ref{sec:dgp}.
We analyze and rank the performance of different causal inference methods using samples (of size N=2500 and 10-dimensional pretreatment covariate space) drawn from (1) the true DGPs and (2) the DGPs learned by training \textit{Credence} on a single sample data set from true DGP with only observed covariates. The goal is to understand if the ranking produced by both approaches are comparable. 

\begin{figure*}
    \centering
    \includegraphics[width=0.49\textwidth]{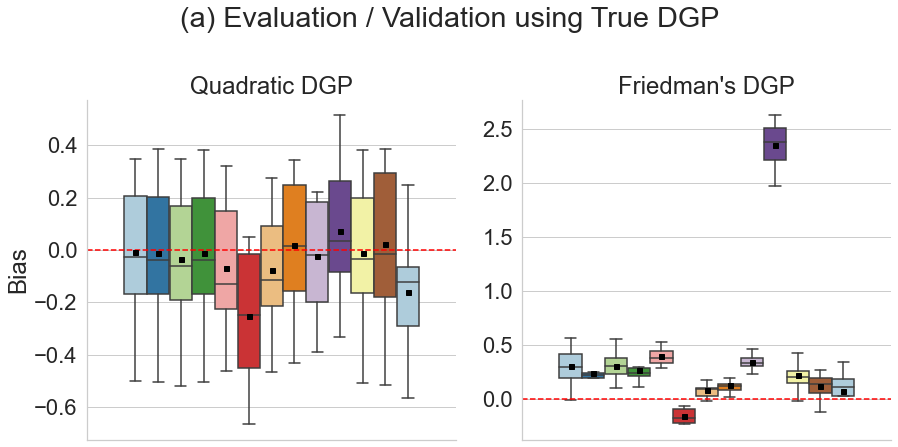}
    \includegraphics[width=0.49\textwidth]{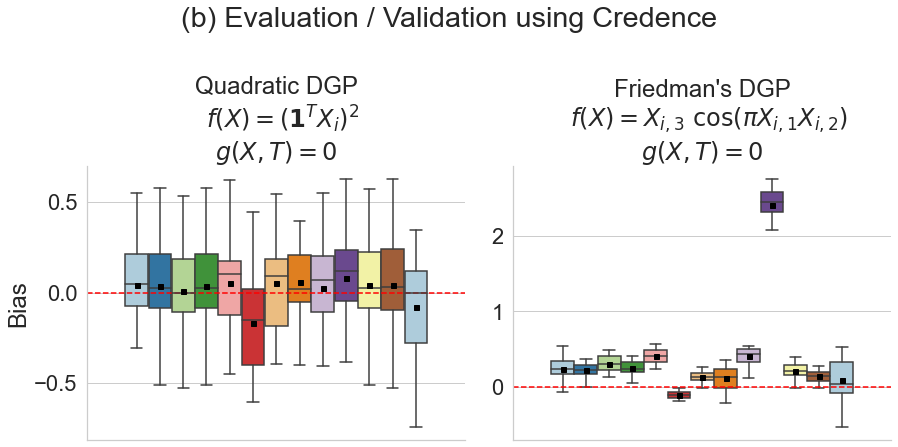}\\
    \includegraphics[width=0.49\textwidth]{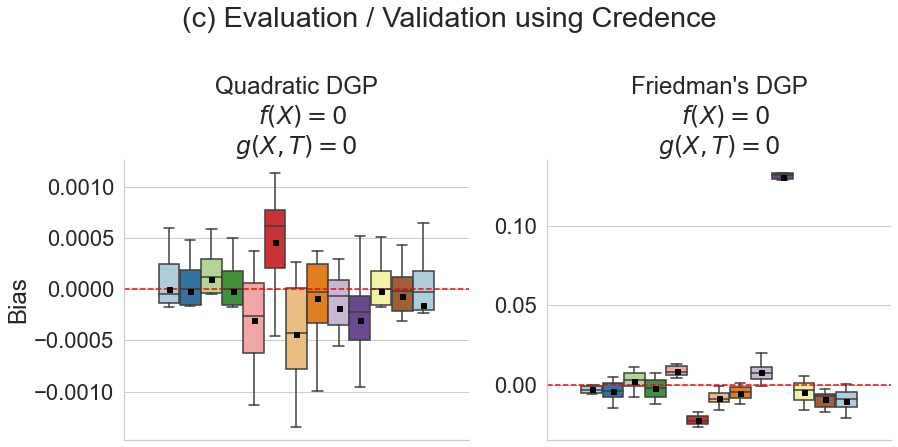}
    \includegraphics[width=0.49\textwidth]{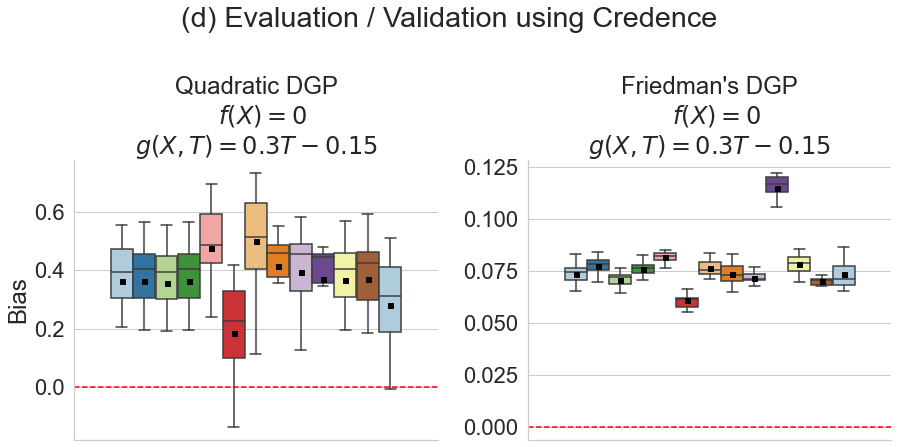}\\
    \includegraphics[width=0.76\textwidth]{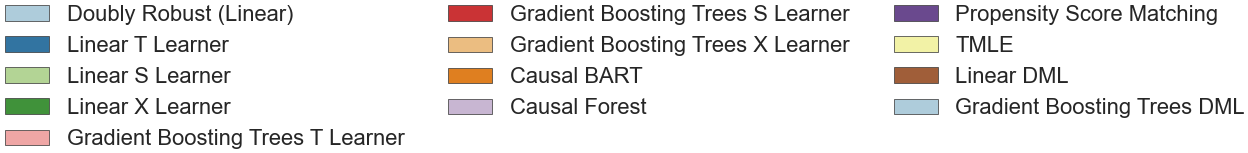}
    \caption{Comparing the performance of various causal inference methods on data generated using (a) \textit{ground truth} quadratic and Friedman's DGPs; (b) Credence trained on a sample of quadratic DGP with $f(X) = (\beta^T X_i)^2 - \beta^T X_i + \alpha^T X_i$  and Credence trained on a sample of Friedman's DGP with $f(X) = X_{i,3}~\cos(\pi X_{i,1} X_{i,2})$ and $g(X,T)=0$ for both cases; (c) Credence trained on a sample of quadratic DGP and Credence trained on a sample of Friedman's DGP with $f(X) = 0$, $g(X,T)=0$; (d) Credence trained on a sample of quadratic DGP and Credence trained on a sample of Friedman's DGP with $f(X) = 0$, $g(X,T)=0.3T-0.15$ for both cases.}
    \label{fig:synth_data}
\end{figure*}

In this study, we train three Credence models for each of the two datasets with different configurations of the treatment effect function ($f(X)$) and selection bias function ($g(X,T)$). (1) For the first one, we shrink $g(X,T)$ towards zero and $f(X)$ towards $(\mathbf{1}^T X_i)^2$ for quadratic DGP and $X_{i,3}~\cos(\pi X_{i,1} X_{i,2})$ for Friedman's DGP (Figure~\ref{fig:synth_data}(b)). (2) For the second one, we constraint both $f(X)$ and $g(X,T)$ to be equal to zero for all $X$ and $T$. (3) Lastly, for the third one, we shrink both $f(X)$ towards zero but constraint $g(X,T) = 0.15(2T-1)$ to understand the sensitivity of different methods to unobserved confounding. For each of the cases, the potential outcomes will choose the values that will minimize Credence's training loss while respecting the above mentioned constraints.

For \textit{Quadratic DGP}, we observe that while most of the methods have comparable accuracy, gradient boosting trees (GBT) S~learner's estimates are the most biased on average (Figure~\ref{fig:synth_data}(a)). Further, linear S~learner and Causal BART and linear DML tend to perform the best. An analogous conclusion can be drawn from the performance assessed using Credence's framework (see Figure~\ref{fig:synth_data}(b) and (c)) where GBT S~learner has the largest bias on average. In Figure~\ref{fig:synth_data}(d), we study the case with no treatment effect but high selection bias that selects units with higher $Y(1)$ into treated group and units with higher $Y(0)$ into control group. There, as expected, we observe that almost all methods are biased. However, the true treatment effect is in 95 percent confidence interval for GBT S~learner and GBT DML.
Further, comparing the performance of causal methods on highly non-linear \textit{Friedman's DGP}, we observe that propensity score matching is highly biased while Causal BART, GBT DML and GBT S~learners tend to be the least biased, as indicated using the true DGP (Figure~\ref{fig:synth_data}(a)). Similar performance is observed when evaluated using Credence-learned-DGPs (see Figure~\ref{fig:synth_data}(b), (c) and (d)). 

The main takeaway from this analysis is that Credence is able to reproduce rankings obtained by an oracle with access to the true DGP in cases where the constraints broadly align with the structure of true DGP. This highlights that the performances evaluated using Credence can provide reliable inferences in such a setting.  


\subsection{Real Data Case Studies}\label{sec:realdata}
The case studies described in this section aim at (1) showcasing the applicability of \textit{Credence} on real world data, and (2) highlight the strength of \textit{Credence} in evaluating causal methods. To address the second objective, we selected data sets that include both experimental and  observational arms. We use the difference of means between treatment and control arms estimated in experimental sample as ground truth ATE and evaluate estimation of ATE in the observational arm against the former. Next, we train \textit{Credence} models only using the observational sample for each of the data sets and evaluate the methods on samples drawn from the trained models. We compare these rankings to determine the extent to which \textit{Credence} performance evaluation of various causal inference methods can be reliable.
\subsubsection{Data Description}
\textbf{Lalonde Temporary Employment Program. }
The National Support Work Demonstration (NSW) temporary employment program is a randomized controlled trial aimed at studying the effect of a temporary employment program in the US on post intervention income level of the participants \citep{Lalonde}. Because this data set is from an RCT, treatment assignment is random and there is no unobserved confounding. An observational counterpart of this data set is the  Population Survey of Income Dynamics (PSID) control sample (described in \cite{website_dehejia}). Treated units from NSW combined with control units from PSID are frequently used as benchmarks to compare the performance of causal inference methods on real data (REFS). The outcome variable for both NSW and PSID samples is an individual's income in 1978. The observed pre-treatment covariates include the age, education, race, degree and 1975 income of participants. The unbiased point estimate of average treatment effect for NSW sample using difference of means estimator is $\$886.30$. Hence, comparing the estimated treatment effect on the combined data set with the ATE from NSW sample benchmarks the performance of causal inference methods.

\textbf{Project STAR. } Project STAR (Student-Teacher Achievement Ratio) is an experiment primarily designed to study the effect of class-size in the earlier grades on short and long-term student performance \citep{project_star_data,mosteller2014tennessee}. A total of 79 schools in the US were chosen across the state of Tennessee in inner-city (17), urban (8), sub-urban (16) and rural areas (38). The experiment has a single cohort that was studied for four years. This included the students entering kindergarten in 1985 and students that initiated their public schooling in first grade in 1986. For each school of the 79 STAR schools, the students and teachers were randomly assigned to one of three treatment arms: a small class (13 to 17~students), a regular-sized class (22 to 25~students) and a regular-sized class with a full-time teacher aide. For our analysis, we consider the students in small class room as treated while all the other students as control. The outcome was measured using a standardized achievement tests in the spring of each year from 1986 to 1989.  Apart from the STAR school data, the information about the non-randomized comparison group was collected concurrently. The comparison group includes information on 1780~students across grades 1 to 3 from 21 schools. The comparison schools were also selected from the same 13 districts as STAR schools. The unbiased point estimate of average treatment effect on Project STAR's experimental sample is $7.24$.

\subsubsection{Analysis} 
\begin{figure*}
    \centering
    \includegraphics[width=0.49\textwidth]{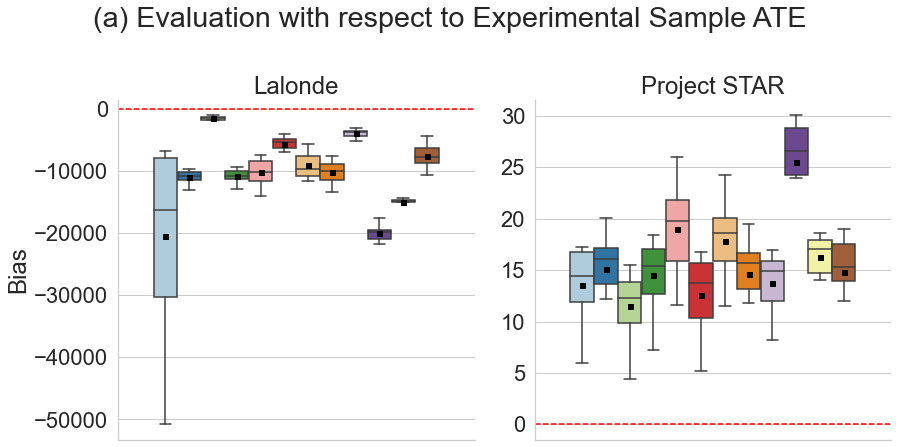}
    \includegraphics[width=0.49\textwidth]{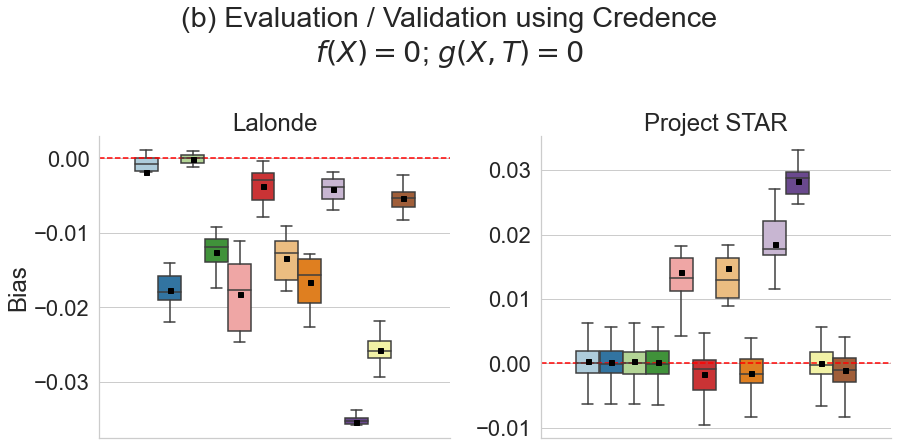}\\
    \includegraphics[width=0.76\textwidth]{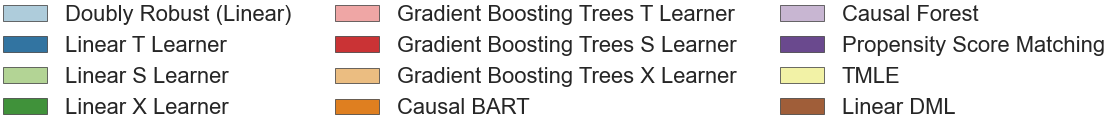}
    \caption{Comparing the performance of various causal inference methods in estimating observation ATE with respect to difference in means experimental ATE estimate for (a) Lalonde NSW+PSID data (b) Project STAR's comparison group data, and with respect to true ATE for data generated using (c) Credence trained on NSW+PSID data (d) Credence trained on Project STAR comparison group data. }
    \label{fig:real_data}
\end{figure*}

\textbf{Analyzing performance with respect to the experimental ATE.} First, we estimated the causal effects of the interventions for Lalonde data and Project STAR's observational data using the various candidate causal inference methods. We compared these estimates with a ``difference of means'' estimate of causal effect in experimental data sets for both studies. As treatment is randomly assigned to units in each of the experimental data sets, the difference of means is an unbiased and consistent estimator. Thus, we treat it as a gold-standard. We compare and rank the performance of ATE estimators (see Figures~\ref{fig:real_data}(a)). For both data sets we find that the bias of each of the estimators on average are very similar. Furthermore, the variance of estimated bias was obtained from 50 bootstrap samples. We observed that the non-parametric DML had significantly larger standard errors on Lalonde data compared to all the other methods, conditional on the observed sample. For Project STAR data, the propensity score method had slightly higher bias relative to other methods.

\textbf{Analyzing performance on Credence generated data.} Next, we trained \textit{Credence} only using the observational component of each study setting $f(X)=g(X,T)=0, \forall X,T$. We sampled 50~data sets using  trained Credence models for Lalonde and Project STAR observational data, respectively. For each sampled data set, we evaluated bias and variance of each candidate ATE estimator. For the Lalonde study, we observed that S-learners, causal forest, linear DML and doubly robust estimators gave the lowest average bias (Figure~\ref{fig:real_data}(b)). For the STAR study, most of the estimators (except GBT T~learner, GBT X~learner, causal forest and propensity score matching) had negligible bias according to evaluation using Credence generated data (Figure~\ref{fig:real_data}(b)). 

For Lalonde's data, rankings based on comparing observational ATE with experimental ATE are largely similar to rankings produced using Credence except with respect to estimated variance of estimators . However, Figure~\ref{fig:real_data}(b) shows that, for Project STAR data, the estimated treatment effect based on observational data is significantly different from experimental data which possibly indicates that the experimental sample lacks external validity \cite{von2018does,justman2018randomized}. Acknowledging this caveat, most methods perform similarly as shown in Figure~\ref{fig:real_data}(a)\&(b). The congruence between both rankings further highlights the validity of \textit{Credence} as an evaluation method anchored at a single observational sample.

\vspace{-3mm}
\section{Discussion and Conclusion}\label{sec:discuss}

\textbf{Implications.}
In this paper, we demonstrate how one can use deep generative models (in this case variational autoencoders) to generate synthetic data sets that are virtually stochastically indistinguishable from a real data set of interest. Importantly, we establish that synthetic data generated via Credence with user-specified treatment effects and hidden confounding bias functions  can accurately recover an oracle's evaluation of various causal effect estimators with knowledge of true data generating mechanism. Researchers continuously develop new causal inference methods with the aim of improving validity of causal effect estimates; however, the task of determining which methods may perform well for data at hand remains complex and time consuming. Therefore, Credence has the potential to become a useful tool to assist the applied researcher in selecting an  adequate method for their observed data. 

\textbf{Selecting $f$ and $g$.} For evaluating causal methods, the best one can hope is to understand the performance of these methods on the synthetic data generated by assuming one or both of the treatment effect function, $f(\cdot)$ and/or selection bias function, $g(\cdot)$. Here, the primary purpose of Credence is not to infer the underlying causal effect but rather assess these methods under certain assumptions about the DGP and rank these methods. As we show in the synthetic data experiments, the suggested approach is to evaluate the methods using \textit{Credence} for a suite of different $f(\cdot)$'s and $g(\cdot)$'s. We recommend two primary strategies for selecting $f$ and $g$. The first approach follows from using the observed data to estimate the larger feasible omitted variable bias possible if one of the observed covariates was omitted. This estimate can help a research understand the potential scale of selection bias ($g$) they can choose. Thus, by doing this one can fixing $g$ to the largest observed omitted variable bias and $\alpha=0$ -- this will let the model learn flexible treatment effects based on the observed data. The second approach is to choose a class of $f$ and $g$, e.g. polynomials of degree 3 or less. One perform a \textit{min-max} search i.e. to find the most adversarial $f$ and $g$ from the class, and choose a causal inference method performs best in that setting.

\textbf{Limitations and Future Direction.}
Researchers face a wide variety of causal measurement problems with different data structures (e.g., small/large number of observations, i.i.d. or time series, small/large number of features). Even when problems share a common structure, underlying data sets can be very unique: some may be sparse, some may be long-tailed, some may have very little feature overlap between treatment and control arms, etc. Although Credence leverages a deep generative model which can flexibly adapt to most of these settings, it still requires careful hyper-parameter tuning for each data set for optimal performance. Hence, a future important direction would involve further developing algorithms to automatically tune hyper-parameters in order to produce synthetic data of the best quality requiring minimal discretion from the applied researcher.

Potential future use for Credence is to perform inference in a manner analogous to the Bootstrap by directly leveraging the learned DGP to obtain repeated samples from Credence. For validity, this would require the researcher to make a non-testable assumption that Credence has successfully uncovered the true underlying DGP for the observed sample. This assumption is more restrictive than required to use Credence to benchmark different causal models. In this latter case, Credence only needs to learn a DGP that can generate data sufficiently similar to the single observed sample, and can reproduce to the extent possible, performance rankings of candidate causal inference estimators similar to those of an oracle with knowledge of the DGP. 

Finally, Credence's evaluation is conditional on the assumptions encoded by the users in learning the DGP, and therefore its diagnostics are as good as these assumptions. For instance, Credence assumes no interference, and no measurement error; if this were not true, the expected performance of candidate estimators of the ATE provided by Credence may be invalid. However, Credence is sufficiently flexible that such evaluation can in principle be conducted under identifying conditions that may be varied in a form of sensitivity analysis. 
\section*{Acknowledgements}
We like to thank researchers and professors at Amazon and Duke University for supporting this work as well as giving valuable feedback. In particular, we appreciate the critical inputs from Drs. Kyle Willett, Alexander Volfovsky, Sudeepa Roy and Babak Salimi. During the work on this research, Harsh Parikh was partially supported by the NSF award IIS-1703431 and a graduate fellowship by Amazon.

\bibliographystyle{apalike}
\bibliography{biblio}

\begin{appendices}

\section{Background on Variational Autoencoders (VAE)}\label{sec:vae}
Autoencoders refer to a particular machine learning estimator that aims to learn a lower dimensional representation of the data which provides a one-to-one mapping between the original data and the lower-dimensional representation \cite{goodfellowdl}. Variational autoencoders (VAE) extend this idea to learn a representation of a high-dimensional complex data set as a standard normal distribution in a lower dimensional latent space. This allows sampling from standard normal distributions in the latent space and projecting it to the original high-dimensional space while ensuring that the distributional properties of the projection and the original data are virtually identical \cite{vae}. Typically, a VAE constitutes two parts -- an encoder and a decoder. The encoder takes in the data 
in the original high dimensional space ($S$) and maps it to $\pi$, a lower dimensional latent space. A decoder performs the opposite operation to that of the encoder, mapping a vector $R$ in lower dimensional latent space to vector $S'$ in higher dimensional space of the original data. The learning algorithm includes (1) encoding the data $S$ as a vector in low-dimensional latent space $\pi=[\mu,\Sigma]$, (2) sampling random vector $R$ in the latent space from a multivariate normal distribution with mean $\mu$ and covariance matrix $\Sigma$ derived from the encoded vector, and (3) decoding the sample vector $R$ from the latent space by projecting it into the space of original data. The VAE loss function has two parts: (i) reconstruction loss (similarity of the input and the decoded output), and (ii) KL divergence between the normal distribution $\mathcal{N}(\mu,\Sigma)$ and standard normal distribution $\mathcal{N}(0,1)$. The reconstruction loss enforces similarity between the empirical distributions of the original and generated samples, while KL divergence ensures that the distribution of latent vectors is as close to a standard normal distribution as possible. This ensures that sampling from standard normal distribution and decoding will have distributional congruence with the original data.
\section{Data Generative Procedure for Synthetic Data Experiments}\label{sec:dgp}
\paragraph{Quadratic DGP:} The pre-treatment covariates $X$ are sampled from a multivariate normal distribution $\mathcal{N}(\mu,\Sigma)$, and the potential outcome function and treatment selection function are defined as follows:
\begin{eqnarray*}
    Y_i(0) &=& \beta^T X_i + \epsilon_{0,i} \;\; \text{ where } \epsilon_{0,i} \sim \mathcal{N}(0,1)\\
     Y_i(1) &=& Y^2_i(0) + \alpha^T X_{i} + \epsilon_{1,i} \;\; \text{ where } \epsilon_{1,i} \sim \mathcal{N}(0,1)\\
     \pi_i &=&\text{expit}(\gamma \times (\mathbf{1}^T X_i) )\\
     T_i &\sim& \text{Bernoulli}(\pi_i)
\end{eqnarray*}

\paragraph{Friedman's DGP:}
This DGP \cite{friedman1991multivariate} was first proposed to assess the performance of prediction methods. We augment Friedman's simulation setup to evaluate causal inference methods. The pre-treatment covariates are sampled from the standard uniform distribution. The potential outcome $Y_i(0)$ is defined by Friedman's function \citep{friedman1991multivariate,Chipman10bart:bayesian}. The expected treatment effect we study is equal to the cosine of the product of the first two covariates scaled by the third covariate.
\begin{eqnarray*}
    Y_i(0) &=& 10~\sin(\pi X_{i,1} X_{i,2}) + 20~(X_{i,3} - 0.5)^2\\&& + 10~X_{i,4} + 5~X_{i,5} + \epsilon_{i,0}\\
     Y_i(1) &=& Y_i(0) + X_{i,3}~\cos(\pi X_{i,1} X_{i,2}) \\
     \pi_i &=& \text{expit}(X_{i,0} + X_{i,1} - 0.5)\\
     T_i &\sim& \text{Bernoulli}(\pi_i)
\end{eqnarray*}
\begin{figure}
    \centering
    \includegraphics[width=0.8\textwidth]{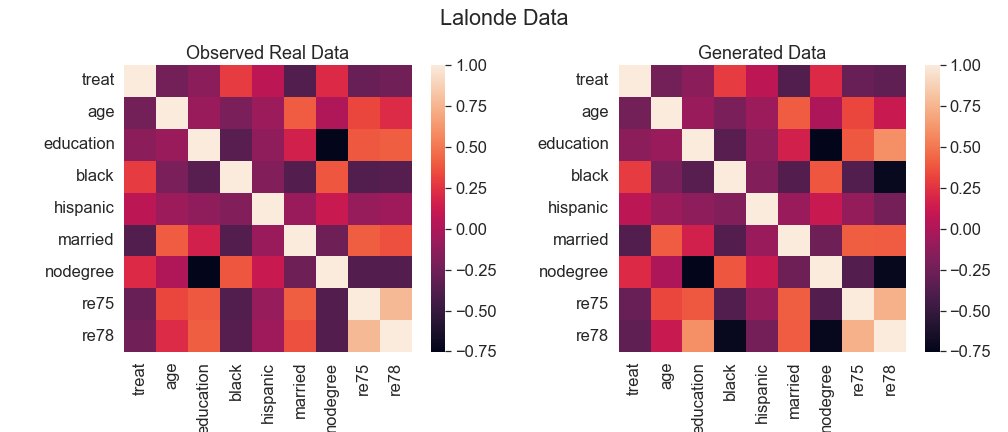}\\
    (a)Correlation Matrix for real and Credence generated Lalonde data\\
    \includegraphics[width=0.97\textwidth]{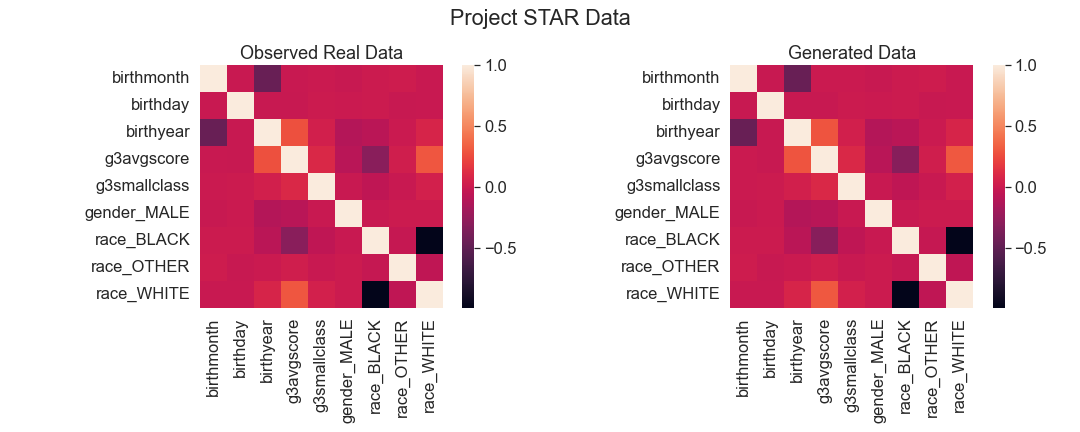}\\
    (b)Correlation Matrix for real and Credence generated Project STAR data
    \caption{Correlation matrices across covariates for comparing the second moments of distributions for the observed real data and Credence generated data for (a) Lalonde and (b) Project STAR. We compare them to understand the distributional similarities between both the DGPs}
    \label{fig:goodness}
\end{figure}
\section{Credence's Goodness of Fit}
In this section, we discuss if the Credence data generated using the learned DGP for Lalonde data and Project STAR data is comparable to the observed real data. We have argued in the paper that the comparing first and second moment is not a sufficient metric of similarity. However, in this section, we compare the correlation matrix of the generated data with the real data because it is easy to visualize and communicate. As we standardize the data removing the mean from each covariate and scaling it by the variance of the same, the vector of means for covariates in the observed and generated data is anchored at 0. In Figure~\ref{fig:goodness}(a) and (b), we show that the correlation matrices (a proxy measure for second moment of the distribution) of the Credence generated data under constraints $f(X)=0$ and $g(X,T)=0$ is visually extremely similar to the correlation matrices of the observables in the real data. This provides a convincing evidence that Credence's generated data has similar distributional properties.
\section{Implementation of Causal Methods}
In this section, we discuss our implementation of causal inference methods studied in Section~\ref{sec:experiment} such as double machine learning (DML), doubly robust estimation, propensity score matching, causal BART, causal forest, metalearners and TMLE. We used the existing libraries and packages 
\begin{itemize}
    \item We used the commonly used MatchIt's implementation of propensity score matching \citep{matchit2011}. We chose the method to match with replacement for estimating ATE.
    \item Our paper uses causal forest implementation from the `grf' R package \cite{grf}. We used the default setting designed by the developer with 2000 number of trees and $\sqrt{p}+20$ variables tried per split.
    \item For causal BART, we use R implementation of BART by Vincent Dorie \citep{dbart}. We only use the method with default hyperparameters. 
    \item Our implementation of GBT DML and linear DML used EconML's implementation of these methods \cite{econml}. We used the scikit-learn's machine learning API for the same \cite{sklearn}. For GBT DML, we used the method with 100 trees, and the linear DML used ridge regression.
    \item Further, we used EconML's implementation of metalearners such as S~learner, T~learner and X~learner \cite{econml}. Similar to DML, we used scikit-learn's ML API to for gradient boosting trees and ridge regression \cite{sklearn}.
    \item We used Paul Zivich's implementation of TMLE \cite{epid}. 
\end{itemize}

\end{appendices}

\end{document}